\pdfoutput=1
\documentclass{article}
\usepackage{arxiv}
\usepackage[T1]{fontenc}
\usepackage[numbers,sort&compress]{natbib}
\usepackage[export]{adjustbox}
\usepackage[utf8]{inputenc}
\usepackage{amsmath,amssymb,mathtools}
\usepackage{graphicx}
\usepackage{booktabs,array,tabularx,longtable,multirow}
\usepackage{algorithm,algorithmic}
\usepackage{float}
\usepackage{enumitem}
\usepackage{placeins}
\usepackage{xcolor}
\usepackage{url}
\setlist{itemsep=2pt,topsep=4pt}
\newcommand{\ThetaA}{\Theta}
\newcommand{\loss}{\ell}
\newcommand{\library}{\mathcal{K}}
\newcommand{\survivors}{\mathcal{S}}
\newcommand{\feasible}{\mathcal{F}}
\newcommand{\suppsec}[1]{Supplementary Section~S#1}
\newcommand{\supptab}[1]{Supplementary Table~S#1}
\newcommand{\suppfig}[1]{Supplementary Figure~S#1}
\newcolumntype{Y}{>{\raggedright\arraybackslash}X}

\AddToHook{env/Abstract/begin}{\righthyphenmin=3\relax}

\usepackage[colorlinks=true,linkcolor=blue,citecolor=blue,urlcolor=blue]{hyperref}
\hypersetup{pdftitle={Consequence-guided clarification: releasing minimum-compliance topology-optimized designs from incomplete specifications},
  pdfauthor={Shaoliang Yang; Jun Wang}}
\title{Consequence-guided clarification:\\
releasing minimum-compliance topology-optimized\\
designs from incomplete specifications}
\author{%
Shaoliang Yang\\
Department of Mechanical Engineering\\
Santa Clara University\\
Santa Clara, CA 95053, USA
\And
Jun Wang\thanks{Corresponding author: \href{mailto:jwang22@scu.edu}{jwang22@scu.edu}}\\
Department of Mechanical Engineering\\
Santa Clara University\\
Santa Clara, CA 95053, USA
}
\date{October 2026}
\renewcommand{\shorttitle}{Consequence-guided clarification}
\begin{document}
\maketitle
\begin{abstract}
A converged topology optimization can solve the wrong problem when
loads, supports, or the domain are underspecified. Assuming defaults, or
asking the problem's author about every detail, ignores how much each
answer changes the design. We propose consequence-guided clarification
(CGC), which bases acceptance on computed mechanical consequences. CGC
computes a reference design for each of a finite list of complete
specifications, called readings. A design is accepted only if its
compliance is at most 5\% above the reference at every reading
consistent with the answers. Otherwise, CGC asks the question whose
worst answer leaves the smallest worst-case excess, and withholds
acceptance at its question limit. To save optimizations, the screening
variant CGC-S postpones some references until a candidate must be
checked. We developed and evaluated both on 36 arXiv-derived tasks,
treating each of 134 readings in turn as intended and simulating correct
answers. In 26 tasks, the default-reading design failed another
reading. Apart from always-asked questions on load direction, sign, and
material budget, asking everything took 188 questions; CGC and CGC-S
each took 105. The acceptance test supplied most of this saving:
applying it to the default design alone took 117. Relative to asking
everything, CGC-S requested 4.6 extra reference designs per
avoided question, each from two optimization starts, counting
optimization time only. References are not certified optima: better ones
pushed three accepted cases above tolerance, and finer meshes changed
some decisions. Within these limits, the test shows which readings a
design covers and which details block acceptance.
\end{abstract}
\section{Introduction}\label{sec:introduction}
Topology optimization determines where material should be placed within a
prescribed domain to meet an objective and constraints
\cite{Bendsoe1989,BendsoeSigmund2003}. The prescribed loads, supports, and
domain are part of the problem: moving a load, changing a support segment,
or keeping a solid region at a load or support can change the material
distribution. A solver may satisfy equilibrium and its volume constraint
for any of these choices, so it cannot tell which one was intended.
Numerical convergence therefore provides little evidence that an
incomplete problem description has been interpreted as its author
intended.

This difficulty arises both when an engineer describes a new problem and
when a researcher reconstructs a published example. A force arrow may
locate a load only approximately. A caption may give a mesh size without
fixing the physical aspect ratio. A support may appear as a symbol whose
extent is unspecified. Benchmark studies state structural optimization
problems explicitly so that solvers can be compared
\cite{Valdez2017,RojasLabanda2015}. An incompletely stated example needs
a further decision before it can serve this way: which complete problems
remain plausible.
We call each unresolved quantity, such as a load position, a \emph{slot}.
We call each complete choice of values for all slots a \emph{reading}.

Two simple responses exist. One asks the problem's author about every
slot. The other commits, without asking, to a \emph{default reading},
usually the one the source appears to show. Neither response considers
how much the answer would change the design, and that varies from slot to
slot.

To measure the effect of a slot, we give each reading a
\emph{reference}, the best design computed for that reading. The
\emph{loss} of a design at a reading is its compliance under that
reading's loads and supports, divided by the reference's compliance,
minus one. Loss is not the change in the optimal
compliance between readings. In a two-spring example
(Section~\ref{sec:problem}), every reading has the same optimal
compliance, yet the design made for one reading performs much worse at
another. A decision to ask or to commit should therefore look at the loss
of the designs it could commit to, not only at the optimal compliance of
each reading.

We treat clarification as a design decision and call the resulting
method consequence-guided clarification (CGC). The method applies established decision rules to specification uncertainty
in topology optimization. The uncertainty concerns the problem specification,
and each regret value must be computed by topology optimization. CGC
takes as input an explicit finite set of readings, written down by the
person reconstructing the problem before CGC runs. Each answer rules out the readings that disagree with it; the
readings that remain are the \emph{open readings}. CGC \emph{releases} a
design, our term for committing to it, only when its loss is within a set
\emph{tolerance}, 5\% here, at every open reading. If none of the designs
computed so far passes, CGC asks about the slot for which, even after
its least favorable answer, the best computed design has the smallest
worst-case loss (Section~\ref{sec:questions}). We do not assign probabilities to
readings; we evaluate the worst reading that remains possible.
Fig.~\ref{fig:method} shows one complete example.

Robust structural optimization chooses one design for multiple scenarios.
Early worst-case design formulations include structural loading families
\cite{KwakHaug1976} and truss designs under specified and occasional
loads \cite{BenTal1997}. Continuum formulations minimize expected
compliance \cite{GuestIgusa2008}, weighted mean and standard deviation
\cite{ZhaoWang2014}, or worst-case compliance \cite{Takezawa2011,Brittain2012}.
Worst-case relative loss instead compares each scenario with its own
best design, so high but unavoidable compliance does not dominate.
Clarification adds the option to remove scenarios by asking which
specification was intended.

Relative loss is a form of regret from statistical decision theory
\cite{Savage1951}. Minimax regret chooses the candidate with the smallest
worst-case regret \cite{KouvelisYu1997,Assavapokee2008}; relative regret
also supports robust numerical-model calibration \cite{Trappler2021}.
Our question rule is closest to regret-based preference elicitation,
which chooses questions to a user by minimax regret \cite{Boutilier2006}.
Here, each alternative is a computed topology-optimized design.

Bayesian active learning selects tests under a known prior over
hypotheses \cite{Golovin2010} and can stop when all consistent hypotheses
lead to the same decision \cite{Javdani2014}. Robust optimization can
also choose which uncertain parameters to observe \cite{Vayanos2026}.
CGC instead assigns no probabilities and accepts a design under a fixed
relative-loss tolerance. This is a satisficing criterion, related to
info-gap decision theory's search for the largest uncertainty set over
which a requirement holds \cite{BenHaim2006}. Our reading set is fixed;
questions reduce it.

Full CGC computes the reference of every reading before asking. Its
screening variant, CGC-S, starts with a smaller design library and delays
further reference optimizations until needed to check a candidate
(Section~\ref{sec:ledger}). A robust design, optimized for all readings
together, can also be tried before asking (Section~\ref{sec:costmodel}).
These references are the best designs found. Numerical controls test how
better references and finer meshes affect release decisions
(Section~\ref{sec:robustness}).

Language-model agents provide a potential interface to this decision.
They generate finite element code \cite{ALLFEM2026}, set up computational
fluid dynamics cases \cite{FoamAgent2026}, plan computer-aided engineering
workflows \cite{Guo2026}, and map commands to microstructure designs
\cite{Kartashov2025}. Some address incomplete specifications: TO-Master
requires confirmation of visualized boundary conditions \cite{TOMaster2026},
TopOptAgents checks formulations and supplies unspecified numerical settings
\cite{TopOptAgents2026}, and InterOPT asks about unresolved optimization
formulations before solving \cite{InterOPT2026}. Our AutoSiMP pipeline
parses text into complete specifications \cite{src_L44}; a release check
could follow that parse. CGC uses computed mechanical consequences to
test a design commitment and needs no language model. We compare it with
a rule that asks by slot type without optimization
(Section~\ref{sec:release}). That rule represents formulation-based
selection, rather than reproducing any of these systems.

We test CGC on a literature benchmark of 36 tasks, each reconstructed
from one underspecified arXiv example. A \emph{policy} is a rule for asking and
releasing: CGC, CGC-S, or a baseline such as the default reading or asking
about every slot. Because we wrote the reading sets ourselves, the replay
can take each listed reading in turn as the intended one and answer every
question exactly. It simulates an exact answerer and does not test a
human workflow. In literature reconstruction, where the source author
usually cannot be asked, the release check still shows which designs need
no answer and whether one design covers every reading. In 26 of the 36
tasks, the design for the default reading fails at least one other
listed reading (Section~\ref{sec:silent}).

The main result is that most question savings come from the release
check itself. CGC asks 105 questions versus 188 for asking every detail;
checking only the design obtained by defaulting unanswered details asks
117, only 12 more than CGC. These counts exclude questions about load
direction, sign, or material budget, which we always ask because a
usable problem statement must fix them (Section~\ref{sec:mandatory}).
Selecting questions by minimax loss
adds no saving over asking in the slots' predeclared order on these
tasks. The contributions are:
\begin{enumerate}
\item A release check based on design regret that accepts a computed
 design only when its loss is within the tolerance at every open reading
 (Section~\ref{sec:problem}).
\item CGC-S, which delays reference optimizations until a candidate
 must be checked and records the information and computation available
 to each decision (Section~\ref{sec:ledger}).
\item A literature benchmark in which both authors judged independently
 which details stay open and which values they may take; the primary
 version keeps the values both accepted, and a second version keeps
 every value either accepted (Section~\ref{sec:benchmarks}).
\item Exact-answer policy replays that separate the release check,
 candidate designs, and question order, with robust alternatives and
 numerical checks (Section~\ref{sec:results}).
\end{enumerate}

The intended use is a decision aid between problem specification and final
design commitment. It can help a researcher or engineer identify which
unresolved slots the computed designs cannot cover within a stated
tolerance. The present study concerns single-material, isotropic,
linear-elastic minimum-compliance reconstructions with
loads that do not change with the design and a volume bound. Stress limits, nonlinear
response, manufacturing requirements, and human interaction costs are
outside the evaluated scope. Sections~\ref{sec:problem}--\ref{sec:method}
define the decision and method; Sections~\ref{sec:implementation}
and~\ref{sec:benchmarks} describe the implementation and benchmarks.
Section~\ref{sec:results} presents the results, followed by their scope
and implications in Section~\ref{sec:discussion}.

\section{Clarification as a design decision}\label{sec:problem}
The decision has three steps: list the complete specifications still
possible, compare a candidate design with a reference at each, and ask
or accept according to the largest relative excess compliance.
Sections~\ref{sec:questions} and~\ref{sec:ledger} turn this check into
question policies. Table~\ref{tab:glossary} collects the core terms for this check.
Policy-specific terms are defined where they first enter the method or
evaluation.

\begin{table}[htbp]\centering\small
\caption{Core terms for the release check.}\label{tab:glossary}
\begin{tabularx}{\linewidth}{lY}
\toprule
Term & Meaning \\
\midrule
\multicolumn{2}{l}{\emph{Specification and release check}} \\
Slot & An unresolved quantity of the specification, with a finite set of admissible values \\
Reading & One complete choice of values for all slots; $t\in\ThetaA$ \\
Pool & All admissible readings of a task, $\ThetaA$ \\
Default reading & The reading a silent-default workflow would choose, $t_0$ \\
Open readings & The readings that agree with every answer so far, Eq.~\eqref{eq:survivors} \\
Reference & The best-known computed compliance of a reading, $V_t$ \\
Loss & Excess compliance of a design at a reading relative to its reference, Eq.~\eqref{eq:regret} \\
Tolerance & The largest accepted loss, $\tau$; 0.05 unless stated \\
Release & Commitment to a design whose loss is at most $\tau$ at every open reading \\
\bottomrule
\end{tabularx}
\end{table}

\subsection{Slots, readings, and open readings}
An ambiguity \emph{slot} is an unresolved quantity in the specification,
such as load position or support extent. Each slot has a finite set of
admissible values grounded in the available description. A \emph{reading}
$t\in\ThetaA$ selects one value for each slot; explicitly inadmissible
combinations are excluded from $\ThetaA$. The default reading $t_0$ is the
specification a silent-default workflow would choose. It is a baseline, not an estimate of the intended reading.

A question asks for one slot value. Given answers $a$, the set of
\emph{open readings} is
\begin{equation}
\survivors(a)=\{t\in\ThetaA:t_s=a_s\text{ for every answered slot }s\}.
\label{eq:survivors}
\end{equation}
The replay treats each admissible reading in turn as the intended reading
$t^\dagger$ and returns exact answers. This produces one case per reading.
The policy sees the answers, not the identity of $t^\dagger$ or the offline
evaluation table.

\subsection{Compliance, feasibility, and computed regret}
For reading $t$, a physical density field $d$ has compliance
\begin{equation}
C_t(d)=\sum_{j=1}^{n_t}w_{tj}f_{tj}^{\mathsf T}u_{tj},
\qquad K_t(d)u_{tj}=f_{tj},\qquad w_{tj}>0.
\label{eq:compliance}
\end{equation}
The positive weights $w_{tj}$ specify the relative contributions of the
independent load cases; each case has its own displacement solution. A design
is feasible for $t$ if its volume over that reading's active domain obeys
the specified bound, its required passive solid elements have density at
least 0.99, and it meets the cross-radius convention described below.
An infeasible design is assigned $C_t(d)=+\infty$ for decision making.

Each reading is optimized from two initial fields. Their lower-compliance
design enters an offline design set $\mathcal D$. Cross-evaluation may
find that a design made for another reading performs better. The
\emph{best-known reference} is therefore
\begin{equation}
V_t=\min\left\{\widehat C_t,
\min_{e\in\mathcal D\cap\feasible_t}C_t(e)\right\},
\qquad
\loss(d,t)=\frac{C_t(d)}{V_t}-1,
\label{eq:regret}
\end{equation}
where $\widehat C_t$ is the best compliance returned by the two starts for
$t$ and $\feasible_t$ denotes feasibility under $t$. Because $V_t$ is a
minimum over $\mathcal D$, no design in $\mathcal D$ has a negative loss;
designs outside $\mathcal D$, such as the robust designs, can have small
negative losses, which are retained. These are computed reference values, with no claim of global optimality.
Because $V_t$ can only overestimate the optimal compliance, a computed
loss can only understate the loss against the optimum, so a passed check
is optimistic.

The primary tolerance is $\tau=0.05$ with zero additional allowance:
the test is not relaxed by the measured spread between the two starts.
This tolerance exceeds the largest relative spread between the two starts of any
literature reference; releases whose margin to $\tau$ is smaller than
that spread are examined in Section~\ref{sec:robustness}.
Counts reported without a stated tolerance use this value;
Section~\ref{sec:robustness} repeats the replay at tolerances from 0.01 to
0.20. A release passes the full-table release check when
\begin{equation}
\max_{t\in\survivors}\loss(d,t)\leq\tau.
\label{eq:release}
\end{equation}
A \emph{pool-unsafe release} is a released design that violates
Eq.~\eqref{eq:release} for at least one open reading. The term refers to
the loss tolerance only; it is not a structural safety assessment. A
\emph{selected-truth exceedance} is a released design for which
$\loss(d,t^\dagger)>\tau$. The first tests whether the release is
justified by the information collected; the second tests its performance
at the reading chosen for that replay case. The tolerance is relative to
the best computed design of each reading. It does not test an absolute
requirement, such as a displacement limit; such a requirement would be
checked separately at every open reading.

\subsection{Why the regret of a design matters}
Comparing optimal objective values alone can miss a large commitment
error. Consider two independent springs with material allocations
$x_1+x_2=1$, stiffnesses $k_i=x_i^3$, and forces
$f_1=\theta^2$, $f_2=(1-\theta)^2$. Their compliance is
\begin{equation}
C_\theta(x_1)=\frac{\theta^4}{x_1^3}
 +\frac{(1-\theta)^4}{(1-x_1)^3}.
\label{eq:springs}
\end{equation}
For every $\theta$ between 0 and 1, the optimum is $x_1=\theta$ and
$V_\theta=1$; \suppfig{2} samples $0.2\leq\theta\leq0.8$. Yet the design
committed at $\theta=0.5$ has $C_{0.2}=3.2896$, or 228.96\% excess
compliance. The optimized value is constant while the cost of using the
wrong design is large.

\subsection{Mandatory questions and the meaning of a passed check}\label{sec:mandatory}
Some unknowns require an answer independently of computed regret.
Reversing a single load changes the displacement sign but leaves
compliance and the optimal density field unchanged. Scaling that load
multiplies compliance quadratically. A reversed single load therefore has
zero regret, and the release check alone would never ask about its sign.
We still ask, because a reconstructed problem must state its load for
any use beyond compliance, such as displacements or stresses; the answer
does not change a compliance design. An unstated volume fraction is also
asked. It could instead be treated like a domain slot, through the
feasibility convention below; we ask it because it sets the material
budget of every design. An unstated direction is treated as mandatory
rather than inferred from a default.
Direction and volume fraction are thus mandatory by choice: compliance
could rank their values, but the replay does not rely on such a ranking.
The replay is conditional on representative mandatory answers
(Section~\ref{sec:benchmarks}). Mandatory questions are counted
separately from questions about varying non-mandatory slots.

Designs are stored on a common grid, mapped to each reading's active
domain, and set to void where their originating domain has no element.
Volume feasibility is checked over the destination domain with relative
tolerance $10^{-6}$. A design generated with a smaller filter radius is
incompatible with a reading requiring a larger radius. This prevents accepting a design generated with less filtering than the
destination reading requires. It is a cross-radius compatibility convention; the geometric minimum feature size
of the final design is not verified. A rank test on constrained rigid-body
modes rejects under-supported readings.

Throughout this paper, a passed release check means that the released
design satisfies the loss tolerance at every open enumerated reading,
relative to the references available to the check and the stated
feasibility convention. Section~\ref{sec:ledger} distinguishes the local
references available to CGC-S from the full table used to evaluate its
releases. Neither check covers an omitted reading, a continuous interval
between sampled values, continuum discretization error, or
manufacturability.

\section{Consequence-guided clarification}\label{sec:method}
\begin{figure}[t]
\centering
\includegraphics[max width=\linewidth]{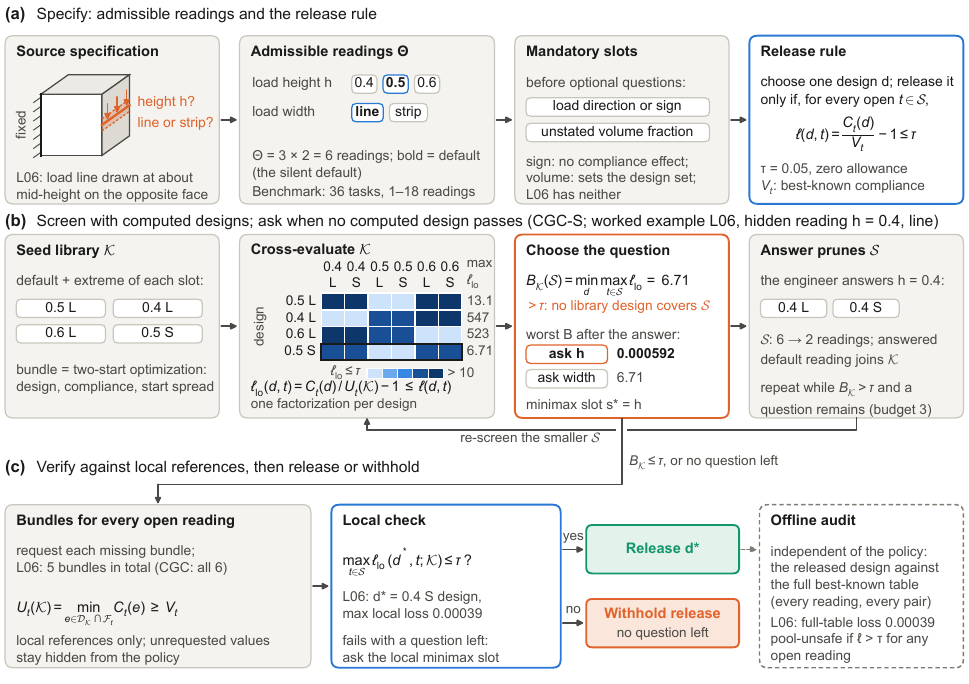}
\caption{CGC-S with zero allowance, with task L06 as a worked example
(hidden reading $h=0.4$, line load). (a)~The source leaves the load
height and width open, giving six admissible readings; a design is
released only if its loss is at most $\tau$ at every open reading.
The library $\library$ holds the readings with computed bundles and
$\mathcal D_\library$ their designs; $U_t(\library)$ is the local reference
of reading $t$, and $B_\library$ is the smallest worst-case local loss of a
library design (Section~\ref{sec:bound}).
(b)~Four seed designs are cross-evaluated against local references. No
library design covers all six readings ($B_\library=6.71$). Asking the
load height leaves a worst-case $B_\library$ of $5.9\times10^{-4}$,
whereas asking the width leaves 6.71, so the height is asked. (c)~The
open reading without a seed bundle receives one, five in total. The best library
design passes the local check with maximum loss $3.9\times10^{-4}$ and is
released. A failed check with no question remaining withholds the release.
The offline audit, which the policy cannot see, evaluates each release
against the full best-known table. CGC: consequence-guided clarification; CGC-S: its screening variant.}
\label{fig:method}
\end{figure}

Fig.~\ref{fig:method} follows one task, L06, through CGC-S; the notation is
defined below. Section~\ref{sec:ledger} states how the local check of
CGC-S relates to the full table.

\subsection{Design libraries and cross-evaluation}
The method maintains a library $\library$ of readings whose reference
bundles have been computed, together with their selected designs
$\mathcal D_\library$. A \emph{reference bundle} is the two-start
optimization of one reading and includes its selected design, compliance,
and relative spread between starts. A requested bundle is charged once per replay case, even when the offline
cache already contains the corresponding runs. This counts the work a
new deployment would request, separately from measured wall-clock time.

For a fixed design, readings with the same domain and support signature
share a stiffness matrix. Their load cases can be evaluated with one
factorization followed by multiple right-hand-side solves. A changed
domain or support set requires another factorization. Thus a load reading
can be evaluated without optimizing that reading. The ledger records
which bundles and design--reading cross-evaluations the policy has
requested; unrequested reference values and start spreads are unavailable
to it.

\subsection{A pointwise bound from a partial library}\label{sec:bound}
If the library contains a feasible design for reading $t$, define
\begin{equation}
U_t(\library)=\min_{e\in\mathcal D_\library\cap\feasible_t}C_t(e).
\label{eq:localref}
\end{equation}
Because the library is a subset of the evaluated offline designs,
$U_t(\library)\geq V_t$. For any feasible candidate with positive reference
compliance, this gives
\begin{equation}
\loss_{\mathrm{lo}}(d,t;\library)
 =\frac{C_t(d)}{U_t(\library)}-1
 \leq\frac{C_t(d)}{V_t}-1=\loss(d,t).
\label{eq:lowerbound}
\end{equation}
If the left-hand side exceeds $\tau$, that particular design fails the
full-reference tolerance at that reading. An infeasible design is rejected
directly; a reading without a feasible library design cannot support a
release through this bound.

This bound rejects a particular design; it does not prove that a question
is necessary, because a different computed or uncomputed design may pass. The
library statistic
\begin{equation}
B_\library(\survivors)=
 \min_{d\in\mathcal D_\library}\max_{t\in\survivors}
 \loss_{\mathrm{lo}}(d,t;\library)
\label{eq:libraryminimax}
\end{equation}
also restricts the candidate designs. It is therefore not generally a
lower bound on the minimax loss over a larger design set. For example,
two library designs with compliance vectors $(1,10)$ and $(10,1)$ have
worst-case loss 9 when both references are 1. Another feasible design with
vector $(1,1)$ would cover both readings without a question. Screening
rules out releases among the computed candidates; its use for choosing
questions is a heuristic search decision.

\subsection{Greedy minimax question choice}\label{sec:questions}
For a candidate set $\mathcal D_c$ and a loss function $L$, write
\begin{equation}
B(\survivors;\mathcal D_c,L)=
 \min_{d\in\mathcal D_c}\max_{t\in\survivors}L(d,t).
\label{eq:minimax}
\end{equation}
If a candidate passes the current check, no non-mandatory question is
needed. Otherwise, score each question by considering every possible
answer. For each answer, retain its consistent readings and find the
candidate with the smallest maximum loss over them. The question's score
is the largest of these values across its answers. Ask the question with
the smallest score:
\begin{equation}
s^*=\arg\min_{s\text{ unasked}}
 \max_{v\in A_s(\survivors)}
 B\bigl(\{t\in\survivors:t_s=v\};\mathcal D_c,L\bigr),
\label{eq:question}
\end{equation}
where $A_s(\survivors)$ is the set of values of slot $s$ among the open
readings. Slots with one such value are skipped. Ties are broken
deterministically: among equal slots, the one whose name sorts first;
among equal designs, the one whose reading label sorts first. The minimum
in Eq.~\eqref{eq:minimax} runs over the finite candidate set, not over all
admissible designs, so a design outside $\mathcal D_c$ may cover readings
that $B$ reports as uncovered. The rule is one-step greedy: it does not search all
future question sequences, and no optimality or approximation guarantee is
claimed for it.

Full-table CGC requests a bundle for every reading before choosing a
non-mandatory question. It uses all computed designs as release
candidates, including designs originating from readings later eliminated
by answers. Its loss is Eq.~\eqref{eq:regret}, and its stopping rule is
Eq.~\eqref{eq:release}. If no design passes within the question budget, it
withholds release.

\subsection{CGC-S with an information ledger}\label{sec:ledger}
CGC-S initially asks the mandatory questions. It then seeds its library
with the answered default reading and the two extreme values of each
varying slot, holding other slots at default values. Duplicate or
inadmissible seeds are removed. It cross-evaluates only designs already
in the library and uses Eqs.~\eqref{eq:libraryminimax}
and~\eqref{eq:question} to select questions. After an answer, the answered
default reading is added when admissible. These steps can prune the set of
open readings before their reference bundles are requested.

When the current library permits a candidate within the tolerance, or no
screening question can be asked, CGC-S requests a reference bundle for
every open reading. Those designs join the library, the local
references $U_t(\library)$ are recomputed, and the best library design
is checked at every open reading. A passing design is released. A failed check
causes another minimax question when one remains within the budget;
otherwise the policy withholds the release. Algorithm~\ref{alg:cgcs} and
Fig.~\ref{fig:method} summarize the procedure. The primary budget is three
non-mandatory questions.

\begin{algorithm}[t]
\caption{CGC-S with zero allowance. CGC: consequence-guided clarification; CGC-S: its screening variant.}\label{alg:cgcs}
\begin{algorithmic}[1]
\REQUIRE Finite readings $\ThetaA$, default reading $t_0$, tolerance
$\tau$, non-mandatory question budget $b$
\STATE Ask mandatory slots; form $\survivors$ using Eq.~\eqref{eq:survivors}
\STATE Request default and slot-extreme seed bundles; initialize $\library$
\LOOP
\STATE Cross-evaluate library designs at open readings; compute $B_\library$
\IF{$B_\library>\tau$ and an unasked varying slot fits the budget}
\STATE Ask the slot from Eq.~\eqref{eq:question} using the library loss
\STATE Update $\survivors$; request the answered default bundle if new
\ELSE
\STATE Request all missing open-reading bundles; update $\library$ and $U_t$
\STATE Recompute the best library design and its maximum local loss
\IF{maximum local loss $\leq\tau$}
\STATE Release that design and stop
\ELSIF{an unasked varying slot fits the budget}
\STATE Ask the local minimax slot; update $\survivors$
\ELSE
\STATE Withhold release and stop
\ENDIF
\ENDIF
\ENDLOOP
\end{algorithmic}
\end{algorithm}

The local and offline checks are distinct. Even after every open reading
has been optimized, a design from an eliminated, uncomputed reading could
lower the full-table reference of an open reading. A passed local check
therefore does not prove Eq.~\eqref{eq:release} against the complete
offline table (\suppsec{10} gives the exact relation between the two
losses). Every reported release is evaluated separately against that
table, without giving its unavailable entries to the policy. The zero
pool-unsafe counts reported below are observed outcomes of this audit,
not a theorem that CGC-S must reproduce full-table decisions on arbitrary
tasks; Section~\ref{sec:robustness} reports how often the local and
full-table references differed.

\subsection{Cost and robust alternatives}\label{sec:costmodel}
The bundle cost of CGC is $|\ThetaA|$. The cost of CGC-S is the number of
distinct readings whose bundles it requests: seeds, any answered default
additions, and the open readings at each verification step, with duplicates
charged once. In addition, it pays for cross-evaluating computed designs.
We report the number of requested design--reading evaluations; each uses
the needed right-hand-side solves, with factorization reuse within a
support group. These counts do not include human answering time;
Section~\ref{sec:screening} reports measured wall-clock times.

Savings require pruning. If all readings stay open because the ambiguity is
harmless, this implementation requests all their bundles before release.
The certified lower bound of Section~\ref{sec:verification} is too weak
in most tasks to support a different stopping test. CGC-S trades computation
against interaction; asking every slot can identify a reading and then
optimize it with only one bundle.

As a no-question alternative, we optimize a reference-normalized
$p$-norm with $p=16$:
\begin{equation}
\min_d\left[\sum_{t\in\ThetaA}
 \left(\frac{C_t(d)}{V_t}\right)^{16}\right]^{1/16}.
\label{eq:robust}
\end{equation}
This baseline is applied only when readings share their domain, filter
radius, and passive regions. The robust optimization starts from the
uniform field, uses the primary filter, projection, continuation, and
optimality-criteria settings, and imposes the smallest volume fraction
among the readings, so that its design is feasible at all of them. It
requires all reading references and is charged as one additional bundle,
$|\ThetaA|+1$ per case. A robust-first policy releases this
candidate when it covers the pool and otherwise uses full CGC, sharing
the references. A minimizer of this norm need not minimize the largest
ratio, and failure of this one optimization does not show that no
covering design exists. Where the $p$-norm design does not cover the
pool, we therefore also solve the min--max problem in bound form,
\begin{equation}
\min_{d,z}\ z\quad\text{subject to}\quad \frac{C_t(d)}{V_t}\leq z
\quad\text{for every } t\in\ThetaA,
\label{eq:minmax}
\end{equation}
with the same filter, projection, continuation, and volume bound, by the
method of moving asymptotes \cite{Svanberg1987} from three starts
(\suppsec{10}). This method is also local, so a bound-form design that
fails leaves open whether a covering design exists. A better robust
design would cover at least the same pools, so robust-first CGC with it
would need no more questions.
Table~\ref{tab:policies} reports the robust variants at zero allowance;
\supptab{18} adds the measured two-start spread as an allowance.

\section{Implementation and numerical verification}\label{sec:implementation}
\subsection{Finite elements, filtering, and optimization}
The solver uses solid isotropic material with penalization (SIMP)
\cite{Bendsoe1989}, four-node quadrilateral elements in plane stress for
two-dimensional problems, and eight-node hexahedra for three-dimensional
problems. It follows the density-filter and projection approach used in
standard density-based implementations
\cite{Bourdin2001,Guest2004,Wang2011,Andreassen2011,Ferrari2020}.
For element centers $r_e$, $x_i$ denotes the element design density,
$r_{\min}$ the filter radius, and $\beta$ the projection sharpness.
The filtered and physical densities are
\begin{align}
\widetilde x_e&=
 \frac{\sum_i\max(0,r_{\min}-\|r_i-r_e\|)x_i}
 {\sum_i\max(0,r_{\min}-\|r_i-r_e\|)},\\
\bar x_e&=
 \frac{\tanh(\beta\eta)+\tanh[\beta(\widetilde x_e-\eta)]}
 {\tanh(\beta\eta)+\tanh[\beta(1-\eta)]},\qquad \eta=0.5,\\
E_e&=E_{\min}+(E_0-E_{\min})\bar x_e^3.
\label{eq:simp}
\end{align}
Element removal defines holes and cut-outs without retaining their
stiffness or including them in the filter or volume. Passive solid and
void element sets are enforced. The volume constraint is imposed on
physical density over the active domain.

Table~\ref{tab:settings} states the common numerical settings. The stored
optimization records use $E_{\min}/E_0=10^{-6}$. The separate void-stiffness
experiment re-evaluates frozen designs at $10^{-9}$; it does not change the
settings of the primary optimizations. Raised-cosine traction patches are
integrated to nodal forces and normalized to their prescribed resultants.
Independent loads are solved independently rather than superposed as a
single force vector.

\begin{table}[tbp]\centering\small
\caption{Common solver and reference settings. Task-specific exceptions are stated in the text. SIMP: solid isotropic
material with penalization; $f$: prescribed volume fraction.}\label{tab:settings}
\begin{tabularx}{\linewidth}{lY}
\toprule
Quantity & Setting \\
\midrule
Elasticity & $E_0=1$, $\nu=0.3$; plane stress in 2-D \\
SIMP and void stiffness & Exponent 3; $E_{\min}/E_0=10^{-6}$ \\
Filter radius & 2.4 elements in controlled 2-D tracks; 1.5 by default elsewhere; source-specific overrides recorded \\
Projection & $\eta=0.5$; $\beta=1,2,4,8$ for 50 iterations each \\
Final phase & $\beta=8$; at most 200 iterations; move 0.02 (L45: 0.01) \\
Convergence & 20-iteration compliance range / absolute mean $<10^{-3}$ \\
Acceptance & Converged; equilibrium residual $\leq10^{-8}$; volume deviation $\leq10^{-4}f+10^{-6}$ \\
Reference starts & Uniform and 5\% smooth sinusoidal perturbation \\
Primary policy & $\tau=0.05$; zero allowance; three non-mandatory questions \\
\bottomrule
\end{tabularx}
\end{table}

An optimality-criteria update uses move limit 0.1 during continuation;
Table~\ref{tab:settings} lists the projection schedule, the final phase,
and the convergence and acceptance tests. Task L45 uses final move limit
0.01 to remove a small period-two oscillation.

The implementation uses Python, NumPy, and SciPy
\cite{Harris2020,Virtanen2020}. Two-dimensional systems use SuperLU
\cite{Li2005}; the literature three-dimensional references use the
PARDISO solver \cite{Schenk2004} of Intel oneMKL in its symmetric
positive-definite mode. Later
controlled three-dimensional runs and refinements use the NVIDIA cuDSS
sparse direct solver library \cite{cuDSS}, version 0.8, on an RTX 5090
graphics processing unit (GPU) with 32~gigabytes of memory. Direct
factorization supports repeated load solves and numerical parity. The
measured relative differences between GPU and central processing unit
(CPU) solutions in the backend check were at most $3.1\times10^{-13}$. These checks establish agreement for the
tested systems rather than a solver performance claim.
Under the authors' direction, an artificial-intelligence (AI) coding assistant (Claude Code,
Anthropic, running Claude Opus~5.5) helped implement the solver,
experiment, figure, and table scripts. The authors reviewed these scripts
before use. The accompanying code regenerates every reported number from
the stored results.

\subsection{Verification and reference quality}\label{sec:verification}

Derivative, force-scaling, reversal, constrained-rigid-mode, and backend
checks verify the tested solver cases (\suppsec{8}; \supptab{6}). The
acceptance tests in Table~\ref{tab:settings} check convergence,
equilibrium, and volume. They do not certify stationarity or a global
optimum. A projected-gradient residual, which measures departure from
first-order stationarity, exceeds 0.1 in 64 of the 134 selected
references. Section~\ref{sec:robustness} tests re-convergence of those
references; \suppsec{14} gives the residual definition.

The difference between two starts measures observed numerical variation,
not an error bound. The primary comparisons use zero allowance; the
alternative spread allowance relaxes the acceptance test
(\supptab{18}).

A lower bound provides a conservative comparison. For physical densities
between zero and one, linear stiffness is at least as large as the SIMP
cubic stiffness. Minimizing compliance with that linear stiffness over
the same volume bound and passive elements therefore bounds the SIMP
optimum from below. A Frank--Wolfe tangent-plane certificate
\cite{Jaggi2013} supplies a valid bound at each iterate
(\suppsec{11}). Section~\ref{sec:robustness} reports how many releases it
can certify. Numerical acceptance, reference quality, and discretization
are thus separate checks.

\section{Benchmarks and evaluation protocol}\label{sec:benchmarks}
\subsection{Controlled ambiguity tracks}\label{sec:controlled}
The two-dimensional tracks use a cantilever, a full Messerschmitt--B\"olkow--Blohm (MBB) beam, an L-shaped
domain, and a plate
with a circular hole. Each has volume fraction 0.4 and a two-element
passive solid strip at the loaded edge. Three tracks vary the load-patch
center, patch width, or division of a fixed resultant between two patches.
Each such geometry with one varying slot is a \emph{track}. Each
track has nine values and two optimization starts per value. Centers span
0.2--0.8 of the loaded entity's length; patch widths span approximately
0.1--0.3 of that length, with a four-element lower limit; load splits span
0.2--0.8. Complete definitions and cross-table maxima are in \suppsec{8}
and \supptab{5}.

The three-dimensional tracks use a $40\times20\times20$ cantilever and a
$48\times12\times24$ deck at volume fraction 0.15, and an L-bracket at
0.2. They vary load height, transverse position, one-element offsets, and
patch width. These reconstructions have no passive load-introduction
region. A cantilever refinement check increases resolution by factors
1.5 and 2, scaling filter radius and patch dimensions consistently; the
largest grid has 128,000 elements. Refined designs use one start, so this
check tests persistence of the observed pattern rather than reference
convergence.

Matched cantilever controls distinguish dimension from reconstruction
choices. They compare two-dimensional volume fractions 0.15 and 0.40
without a passive strip, 0.40 with the two-element strip, and the
three-dimensional volume-fraction-0.15 track. All evaluate the mid-height
load design when its load is moved. A separate three-slot cantilever
combines load center, patch width, and angle into 18 readings to test
question ordering beyond the literature tasks.

\subsection{Frozen literature frame}\label{sec:literature}
The frame was retrieved from the arXiv application programming interface
(API) on 22 October 2026 with a query for papers that mention topology
optimization together with compliance or stiffness; first submissions
from 1 January 2015 on were retained. The resulting 247 papers were
screened in a hash order fixed at retrieval, before screening began
(\suppsec{1} gives the query, the order, and the screening blocks). The
screening records date from 22 October 2026, and the author adjudication
records were completed on 23 October 2026.
We drew the frame from arXiv rather than from journal databases for two
reasons. Every source text and figure is open, so any reader can check
each encoding against its source. The stated query, date range, and
retrieval time also define the frame exactly. The frame is not limited to
unpublished work: 22 of the 36 tasks have confirmed published versions, and
14 of the 15 versions we compared leave open the same details as the
encoded text (Section~\ref{sec:provenance}). It is not a representative sample of all engineering
specifications or published topology-optimization papers.

\begin{figure}[t]
\centering
\includegraphics[max width=\linewidth]{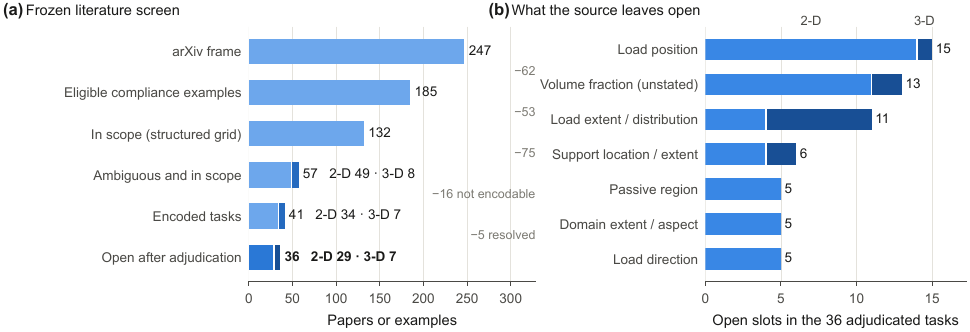}
\caption{Literature benchmark construction. (a) Frozen arXiv frame,
screening, encoding, and adjudication. The ambiguous count precedes author
adjudication. (b) Open slot categories in the 36 adjudicated tasks, separated
by dimension. A task can contribute several categories. Unstated volume
fractions are mandatory questions. A slot with two labels is counted under
its first: the void layout of L11, labeled passive region and domain, is
counted as a passive region here, whereas the slot-type rule of
Section~\ref{sec:release} labels it domain (\suppsec{11}).}
\label{fig:census}
\end{figure}

The eligible primary example had to concern structural compliance
or stiffness. Reconstruction scope further required a structured rectangle
or box, possibly with removed elements or passive regions, isotropic
linear elasticity, and design-independent point, line, or patch loads.
The source optimization parameterization could differ from SIMP. Of 185
eligible examples, 132 were within this scope; 57 remained ambiguous after
text and figures were considered, including eight three-dimensional
examples (Fig.~\ref{fig:census}).

Of the 57 ambiguous examples, 41 were initially encoded
\cite{src_L01,src_L02,src_L04,src_L05,src_L06,src_L07,src_L08,src_L09,src_L11,src_L12,src_L13,src_L14,src_L15,src_L16,src_L17,src_L18,src_L19,src_L20,src_L21,src_L22,src_L23,src_L24,src_L25,src_L27,src_L28,src_L29,src_L30,src_L31,src_L33,src_L34,src_L35,src_L36,src_L37,src_L38,src_L39,src_L40,src_L41,src_L42,src_L43,src_L44,src_L45};
\supptab{1} maps each source to its task. The other 16 were excluded because
the particular random instance could not be reconstructed as a finite set,
the apparent alternatives were equivalent or resolvable from a cited
dataset, or a required load or filter representation was unavailable.
Every excluded example and reason is in \supptab{4}. Screening spot checks
corrected 3 of 14 records in the second block and 7 of 17 in the third;
the other records were not re-checked, so the frame counts are estimates
with that error rate.

Labels such as L06 are stable encoding identifiers retained to match the
stored records. They are not frame positions or ranks and need not be
consecutive. \supptab{1} supplies each source identifier; \supptab{3}
lists the encodings resolved by adjudication.

\subsection{Annotation, adjudication, and reconstruction}

AI agents using Claude produced screening records and first encodings.
Both authors independently adjudicated all 41 initial encodings and 16
exclusions, with blind model re-annotation as a second opinion. Table~\ref{tab:adjudication}
shows their coverage and decisions: adjudicator A read 32 source texts
directly and judged the other nine from sheet evidence; B read all 41
with figures. Original records are Supplementary Files A and B.

Because B retained every proposed non-mandatory slot, the primary open
or closed decisions follow A. The primary values use intersections of
author-accepted values with two explicit reconciliation exceptions;
the union version retains values accepted by either author
(\suppsec{3}). Adjudication fully resolved five of the 41 encodings, leaving 36 primary tasks, 29 in
two dimensions and seven in three, with 134 readings. These are author
judgments about a finite reconstruction set. Mandatory slots use
representative answers apart from the L05 sign control; alternative
mandatory answers were not exhaustively replayed.

The solver and reconstruction conventions are standardized rather than
reproducing every source method. Grids, admissible values, defaults,
source evidence, and reductions are listed in \supptab{1}; consequences
hold at those reconstruction resolutions. Three encoded and two excluded
papers are the authors' own work; they stayed at their frozen frame
positions. Controls removing these sources and the tasks judged from
sheet evidence are reported in \suppsec{11} and \suppsec{14}. The detailed
adjudication, reconciliation, and reconstruction account is preserved in
\suppsec{14}.

\begin{table}[tbp]\centering\small
\caption{Author adjudication of the 41 initial encodings.}\label{tab:adjudication}
\begin{tabularx}{\linewidth}{lYrr}
\toprule
Adjudicator & Source texts checked directly & \shortstack{Non-mandatory\\slots kept open} & \shortstack{Mandatory\\slots kept} \\
\midrule
A & 32 of 41 tasks & 44 of 69 & 17 \\
B & 41 of 41 tasks, with figures & 69 of 69 & 17 \\
Both & -- & 44, identical values for 40 & 17 \\
\bottomrule
\end{tabularx}
\par\smallskip\raggedright\footnotesize The 69 candidate non-mandatory slots include 17 proposed additions. Mandatory slots are kept by rule (Section~\ref{sec:problem}), so the open or closed decision concerns only non-mandatory slots.
\end{table}

\subsection{Policies, denominators, and evaluation}
The silent default commits to the default design without a question.
A mandatory-only baseline asks mandatory slots and defaults the rest.
Ask-all asks about every open slot and requests one bundle for the
identified reading. CGC and CGC-S use Section~\ref{sec:method}; a fixed-order policy
uses CGC's release check but asks slots in a prescribed order. Two
default-only baselines separate the release check from CGC's candidate
designs. The default check releases the default design of the answers so
far when its loss is within $\tau$ at every open reading, and otherwise
asks the next slot in declared order; its ledger version runs under the
seeds and local references of CGC-S. The feasibility screen needs no
reference: it releases the default design when that design is feasible
at every open reading and otherwise asks every remaining slot
(\suppsec{10}). A hindsight
subset baseline enumerates answered-slot subsets for the known replay
truth and tests the reference designs of the open readings. It compares question
counts; it is neither deployable nor a guarantee of globally minimal
questions.

Every task is replayed once per admissible reading. Case-weighted totals
therefore give more weight to tasks with more readings. We also report
means formed by averaging within each task first. Question totals exclude
mandatory questions unless stated. Bundle counts charge each reading once
per case and include both starts. The offline evaluator checks every
release against all open readings using the exhaustive best-known
references, withholding unavailable entries from the policy ledger.

The policies and ledger were revised using the encoded tasks. There is no
held-out evaluation. \suppsec{12} lists, with dates, which rules were
fixed before the literature replay and which were revised after its
results were seen. Comparisons describe development-set behavior under
exact-answer replay. The exploratory language-model front end is specified
and evaluated separately in \suppsec{6}.

\section{Results}\label{sec:results}
\subsection{The consequence of an ambiguity depends on geometry and load introduction}

\begin{figure}[t]
\centering
\includegraphics[max width=\linewidth]{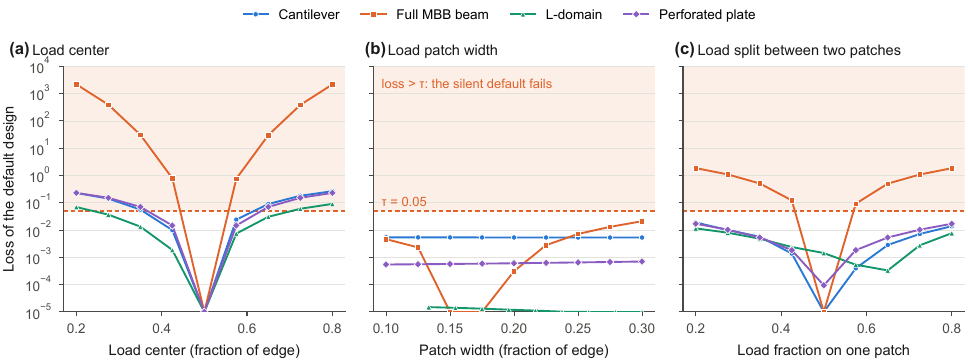}
\caption{Default-design loss in controlled two-dimensional tracks. Each
geometry has a passive solid strip at the loaded edge. Tracks vary load
center, width, or split between patches. The dashed line is $\tau=0.05$,
and shading marks losses above it;
values below $10^{-5}$ are drawn at $10^{-5}$. Default-design curves and
full cross-table maxima are distinct quantities. MBB:
Messerschmitt--B\"olkow--Blohm beam.}
\label{fig:controlled2d}
\end{figure}

Figure~\ref{fig:controlled2d} shows that load center, patch width, and
load split have different consequences across geometries. In three
geometries every patch-width cross-table entry has at most about 0.54\%
loss. The full MBB beam reaches 12.4\% over all width-specific designs
and evaluation widths. Its single default-width design reaches only
2.1\% over the same evaluation widths. Center-position changes can be much more consequential:
the MBB full-table maximum is 2,199 with the passive load strip. It
persists at lower void stiffness (\supptab{5}; \suppsec{14}).

\begin{figure}[t]
\centering
\includegraphics[max width=\linewidth]{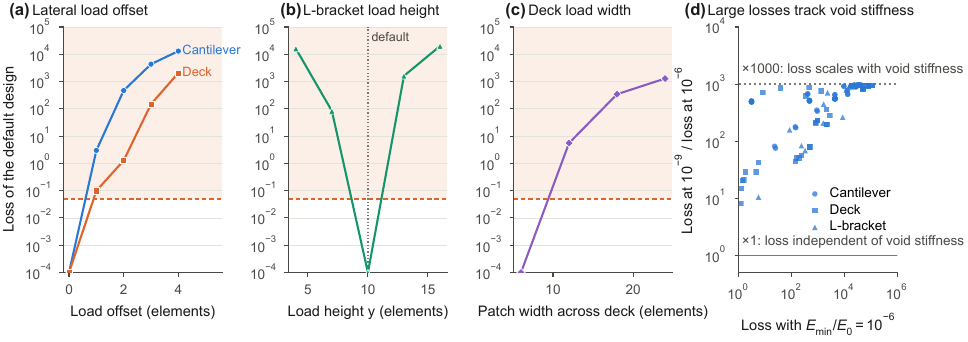}
\caption{Controlled three-dimensional tracks with no passive
load-introduction region. (a--c) Default-design loss under load offsets,
L-bracket load-height changes, and deck patch-width changes; shading
marks losses above $\tau=0.05$. (d) Frozen
designs re-evaluated at two void stiffnesses: ratio of losses at
$E_{\min}/E_0=10^{-9}$ and $10^{-6}$ for entries with baseline loss above
1. Baseline optimizations use $10^{-6}$; $E_{\min}/E_0$ is the
void-to-solid stiffness ratio. These tracks show the convention
without a load-introduction region, not an effect of dimension
(Table~\ref{tab:matched}).}
\label{fig:controlled3d}
\end{figure}

The three-dimensional tracks in Fig.~\ref{fig:controlled3d} have no passive
load-introduction region (a prescribed solid region around the loaded
area) and show large losses under small load changes.
Their magnitudes depend on artificial void stiffness and are not physical
performance percentages. Matched cantilever controls (Table~\ref{tab:matched})
show that the passive loading region, rather than dimension alone, can
change whether a shifted load is tolerated. The cantilever pattern
persists under mesh refinement, while the largest magnitudes remain
regularization-dependent. \suppsec{14} preserves every reported offset,
void-stiffness, and refinement comparison.

\begin{table}[tbp]\centering\small
\caption{Matched cantilever controls: loss of the mid-height design at two shifted loads, at heights $0.35H$ and $0.2H$ with $H$ the cantilever height. Volume $f$: volume fraction.}\label{tab:matched}
\begin{tabularx}{\linewidth}{llYrr}
\toprule
Dimension & Volume $f$ & Passive loading region & $0.35H$ & $0.2H$ \\
\midrule
2-D & 0.15 & None & 718 & 6,108 \\
2-D & 0.40 & None & 2,075 & 17,243 \\
2-D & 0.40 & Two-element strip & 0.042 & 0.143 \\
3-D & 0.15 & None & 400 & 17,665 \\
\bottomrule
\end{tabularx}
\end{table}

\subsection{Silent defaults often exceed the tolerance within the reading pool}\label{sec:silent}

The silent default fails at least one open reading in 26 tasks and
101 of 134 replay cases; it fails the reading selected for the replay in
65 cases. These are the pool-unsafe and selected-truth outcomes defined in
Section~\ref{sec:problem}. These are different outcomes:
a design can fail an open reading while passing the reading selected for
the replay. Because mandatory answers are largely fixed to
representatives, the mandatory-only baseline has the same exceedance
counts. \supptab{2} gives the task-level outcomes.

The failures mix reconstruction effects. Seven tasks fail a domain or
passive-region feasibility rule. Many other two-dimensional failures
place most of a load on numerically void material, and three tasks
include single-node load or support readings. Diagnostic categories
describe the failing readings, rather than establishing their causes
(\supptab{13}). A shared solid region around all possible loaded nodes
makes the default pass in three tasks. Removing single-node readings as
well leaves 20 failing tasks out of 36. Simply filling missing passive
regions can instead violate the volume bound. \supptab{14} and
\suppsec{14} retain the recipes, per-task outcomes, and qualifications.

Most sampled consequences are either below the tolerance or far above
it. No two-dimensional task has a default-design loss above $\tau$ and
at most 10; only four three-dimensional tasks are in that range.
The benchmark therefore tests graded consequences sparsely. Alternative
position ranges and load-introduction conventions are tested in
\suppsec{10}; they do not establish robustness to arbitrary encodings.

\begin{figure}[t]
\centering
\includegraphics[max width=\linewidth]{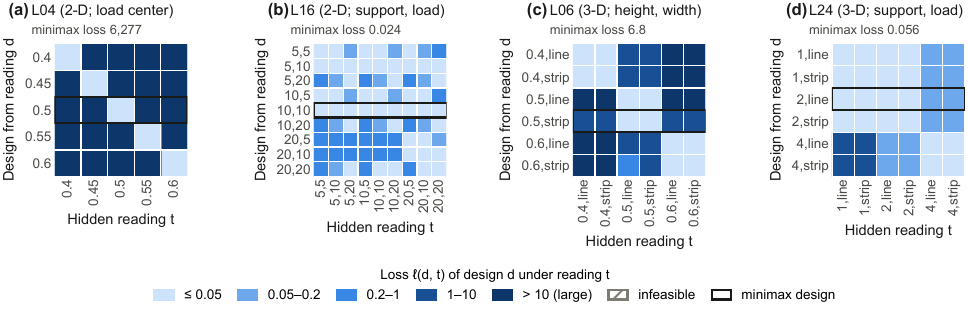}
\caption{Cross-evaluation tables for four literature tasks. Rows identify
the originating design reading; columns identify the evaluation reading.
Colors bin excess compliance relative to the best-known reference;
hatching denotes destination-reading infeasibility. The outlined row is
the design with the smallest worst-case loss, given above each table.
Examples include
load-position, load-distribution, and support-extent ambiguities. In (b),
labels give the support and load lengths in percent of the domain side length $L$.}
\label{fig:crosstables}
\end{figure}

\begin{figure}[t]
\centering
\includegraphics[max width=\linewidth]{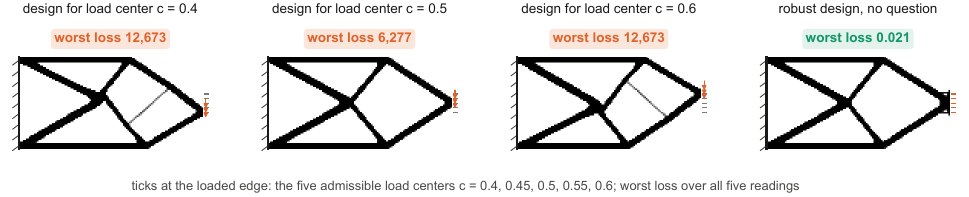}
\caption{L04 designs for three load-center readings and a robust design.
Each worst loss is over all five readings; green labels are at most
$\tau$ and orange labels exceed it. Arrows mark the load center of the
design's own reading, and ticks mark the five admissible centers (all
five for the robust design). Robust coverage applies to that pool; an
omitted load position can still fail (\suppsec{8}). The source also makes
a region around the load solid; this encoding omits that region
(Section~\ref{sec:silent}). In the union encoding, a design with such a
region at one load center is infeasible at every reading that places the
region at another center, so that encoding cannot isolate the region's
effect on these losses; Table~\ref{tab:matched} isolates it in matched
controls.}
\label{fig:gallery2d}
\end{figure}

\begin{figure}[t]
\centering
\includegraphics[max width=\linewidth]{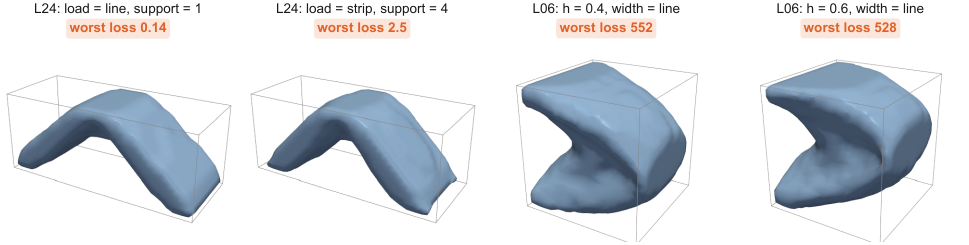}
\caption{Three-dimensional designs. L24 compares line loading with narrow
supports and strip loading with wider supports; L06 compares two load
heights. Each label gives that design's worst pool loss, which can exceed
the default design's worst loss; all four exceed $\tau$. Surfaces use the 0.5 density isovalue. The
loads, supports, and domains of each reading are defined in
\supptab{1}.}
\label{fig:gallery3d}
\end{figure}

Figures~\ref{fig:crosstables}--\ref{fig:gallery3d} connect these comparisons
to the underlying designs. Rows of the cross-evaluation tables identify
a design's originating reading; columns give its destination reading.
In L04 a robust design covers all enumerated load positions, while
position-specific designs can fail other readings. The L24 and L06
examples show distinct three-dimensional designs under different support
or load readings. Neither visual plausibility nor these isosurfaces
supply a separate structural safety assessment.

\subsection{The release check accounts for most of the saving}\label{sec:release}

\begin{figure}[t]
\centering
\includegraphics[max width=\linewidth]{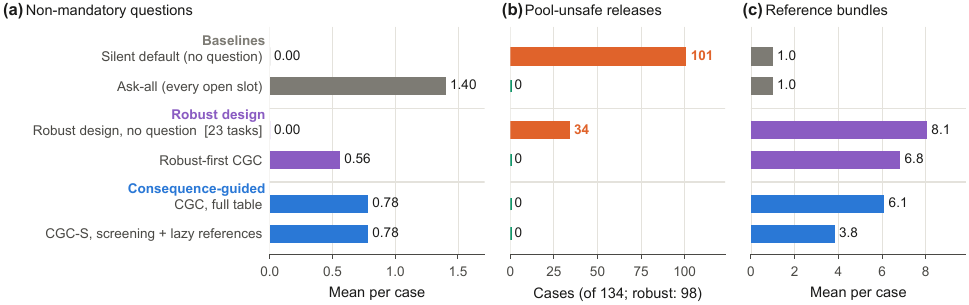}
\caption{Primary benchmark policy outcomes. CGC: consequence-guided
clarification; CGC-S: its screening variant. Question and bundle bars are
means per case; unsafe releases are case counts. All policies use zero
allowance and $\tau=0.05$. Bar colors group baselines, robust policies
and consequence-guided policies; orange marks pool-unsafe releases. The
no-question robust baseline applies to 23 tasks and 98 cases; other
policies apply to 36 tasks and 134 cases.}
\label{fig:policies}
\end{figure}

At zero allowance and $\tau=0.05$, CGC and CGC-S release in all 134 cases
without a pool-unsafe release or selected-truth exceedance
(Table~\ref{tab:policies}; Fig.~\ref{fig:policies}). For full CGC this
holds by construction; for CGC-S it is an observed outcome of the
separate full-table audit. They ask 105
non-mandatory questions versus 188 for ask-all, a 44.1\% reduction.
These are questions, rather than interaction rounds: batching every
question can make ask-all require fewer rounds (\suppsec{13}).

\begin{table}[htbp]\centering\small
\caption{Policy outcomes for 134 primary replay cases at $\tau=0.05$. CGC: consequence-guided clarification; CGC-S: its screening variant.}\label{tab:policies}
\begin{tabularx}{\linewidth}{Yrrrrr}
\toprule
Policy & \shortstack{Non-mandatory\\questions} & Releases & \shortstack{Pool-\\unsafe} & \shortstack{Truth\\exceed.} & \shortstack{Bundles\\per case} \\
\midrule
Silent default & 0 & 134 & 101 & 65 & 1.00 \\
Feasibility screen & 28 & 134 & 81 & 54 & 1.10 \\
Ask-all & 188 & 134 & 0 & 0 & 1.00 \\
Default check & 117 & 134 & 0 & 0 & 6.09 \\
Default check, ledger & 117 & 134 & 0 & 0 & 3.81 \\
CGC & 105 & 134 & 0 & 0 & 6.10 \\
CGC-S & 105 & 134 & 0 & 0 & 3.84 \\
Robust & 0 & 98 & 34 & 33 & 8.06 \\
Robust-first CGC & 75 & 134 & 0 & 0 & 6.84 \\
Type rule, load extent defaulted & 143 & 134 & 15 & 8 & 1.00 \\
\bottomrule
\end{tabularx}
\par\smallskip\raggedright\footnotesize All rows use zero allowance. The robust rows use the $p$-norm design; with the bound-form min--max design where it is better (\supptab{20}), robust alone is pool-unsafe in 32 cases and robust-first CGC asks 73 questions. Robust alone applies to 23 tasks and 98 cases; other rows cover 36 tasks and 134 cases. The default-check rows charge only the readings left open after the mandatory answers; charged like CGC, the first would need 818 bundles, 6.10 per case. The type rule asks by slot type without optimization (Section~\ref{sec:release}); narrower type lists are in \supptab{15}, and the spread-allowance and spread-margin replays in \supptab{18}. The zero pool-unsafe count of full CGC holds by construction; those of ask-all and CGC-S are observed (ask-all has one at $\tau\leq0.02$, Table~\ref{tab:tau}). Pool-unsafe refers to the loss tolerance, not to structural safety.
\end{table}

Most of this saving comes from the acceptance test. Applying it only to
the design obtained by defaulting unanswered details asks 117 questions,
without a pool-unsafe release. That accounts for 71 of the 83 avoided
questions. Other computed candidates supply the remaining 12; those
cases involve passive-region or single-node reconstruction conventions
(\suppsec{14}). The default-only check still needs reference values, so
it does not provide the same saving without computation.

The minimax question rule adds no primary-set saving over asking slots
in their declared order: both ask 105 questions with the same release
check and candidates. Only 11 tasks have two varying non-mandatory slots;
the others have at most one. This limited agreement supplies no general
question-optimality guarantee. \supptab{17} gives every order comparison;
Section~\ref{sec:union} tests richer reading sets.

The reference-free alternatives in Table~\ref{tab:policies}, a feasibility
screen and a rule that asks by slot type, release designs that exceed
the tolerance. A slot's name or feasibility alone therefore does not
reproduce the computed acceptance test on these tasks.

A robust design can avoid additional questions at higher computation
cost. Robust-first CGC asks 75 questions with the $p$-norm candidate and
73 when the better bound-form min--max candidate is used. Robust alone
applies to 23 tasks and 98 cases; its failures do not prove that no
covering design exists, because those optimizations are local
(\supptab{20}). The full baseline definitions, allowances, order
comparisons, and per-task explanations are retained in \suppsec{14}.

\subsection{Most releases hold against better references; decisions near the tolerance depend on the mesh}\label{sec:robustness}

The primary zero-exceedance counts use two-start references, sampled slot
values, and $\tau=0.05$. Full CGC is judged by its own stopping criterion,
so its zero holds by construction. CGC-S uses local references and is
scored separately against the full table. At its 134 releases, local and
full-table references agree at 313 of 314 open readings; the sole gap,
in L11, does not change that release. This benchmark therefore offers
only a weak test of the local-reference approximation.

\subsubsection{Additional starts and mesh refinement}
Nine selected tasks test decision stability under six starts or a mesh
refined by two in each direction (Table~\ref{tab:near}). The
selection criteria, consistent scaling, and frozen-design rescoring
are specified in \suppsec{8} and \suppsec{14}. Additional starts change
decisions only in L25; refinement changes them in L18, L21, and L25.
The changes are not confined to losses just below the tolerance: L21
moves from minimax loss 0.0109 to 0.1326. The other 21 two-dimensional
tasks were refined in the same way and keep every policy decision
(\supptab{16}).

\begin{table}[tbp]\centering\small
\caption{Full-table minimax loss at zero allowance under additional starts and consistent mesh refinement.}\label{tab:near}
\begin{tabularx}{\linewidth}{lrrrrY}
\toprule
Task & Two starts & Six starts & $2\times$ mesh & Readings & Decision stability \\
\midrule
L18 & 0.0481 & 0.0406 & 0.0550 & 3 & Stable for six starts; changes with mesh \\
L24 & 0.0561 & 0.0578 & 0.0642 & 6 & Stable for six starts; stable with mesh \\
L37 & 0.0438 & 0.0440 & 0.0427 & 4 & Stable for six starts; stable with mesh \\
L16 & 0.0239 & 0.0250 & 0.0307 & 9 & Stable for six starts; stable with mesh \\
L25 & 0.510 & 0.492 & 0.504 & 4 & Changes for six starts; changes with mesh \\
L11 & 7.75 & 8.37 & 8.41 & 3 & Stable for six starts; stable with mesh \\
L20 & 0.0166 & 0.0166 & 0.0210 & 2 & Stable for six starts; stable with mesh \\
L21 & 0.0109 & 0.0110 & 0.1326 & 2 & Stable for six starts; changes with mesh \\
L36 & 0.0115 & 0.0111 & 0.0225 & 6 & Stable for six starts; stable with mesh \\
\bottomrule
\end{tabularx}
\par\smallskip\raggedright\footnotesize A decision is stable when every case keeps its question count and release outcome. L25 and L11 have a minimax loss far above $\tau$ but a release decided after an answer, with a margin below the largest two-start spread.
\end{table}

Improved references also test the designs already released, without
changing those designs. A common pool of table, robust, and six-start
designs pushes only the three L18 releases above $\tau$, to 0.0504.
Re-converging the 64 references farthest from stationarity changes none
by more than 1.7\% and pushes the same releases to 0.0507. Replaying CGC
and CGC-S with those references adds one question to each L18 case,
108 questions in total, and releases in all 134 cases without a
pool-unsafe release. A decision can therefore be stable against the
original table and still depend on reference quality or discretization.

\subsubsection{Conservative certification and coverage checks}
A certified replay replaces computed references with the lower bounds
of Section~\ref{sec:verification}. At $\tau=0.05$ and unlimited question
budget, it releases in 37 of 134 cases. On that released subset, CGC asks
40 questions versus 66 for ask-all. Certification preserves a question
saving on fewer releases; the subsets change with tolerance
(\suppsec{13}).

Optimized midpoint readings and held-out values in controlled tracks
found no tolerance exceedance, but neither test covers every
intermediate value (\suppsec{10}; \suppsec{11}). A two-start-spread
margin can also flag numerically unresolved releases; it is a heuristic,
not an error bound. Its outcomes and the full diagnostic account are in
\suppsec{14}.

Table~\ref{tab:tau} varies the tolerance. At 0.01 and 0.02, CGC-S and
ask-all each make one pool-unsafe release in L11 because a design from
an eliminated reading lowers the best-known reference compliance. That
better denominator raises the measured relative loss of the released
design, revealing an exceedance missed by the local check. Full CGC has none by
construction. The observed zero at 0.05 therefore supplies no universal
full-table guarantee for the local check.

\begin{table}[tbp]\centering\small
\caption{Primary benchmark (134 cases) replayed at five tolerances, zero allowance. CGC: consequence-guided clarification; CGC-S: its screening variant.}\label{tab:tau}
\setlength{\tabcolsep}{3.5pt}
\begin{tabular}{lrrrrrrrrrr}
\toprule
$\tau$ & \shortstack{Silent\\unsafe} & \shortstack{Ask-all\\questions} & \shortstack{Ask-all\\unsafe} & \shortstack{CGC\\questions} & \shortstack{CGC\\releases} & \shortstack{CGC-S\\questions} & \shortstack{CGC-S\\releases} & \shortstack{CGC-S\\unsafe} & \shortstack{CGC-S\\bundles} & \shortstack{Robust-first\\questions} \\
\midrule
0.01 & 123 & 188 & 1 & 136 & 134 & 136 & 134 & 1 & 3.62 & 127 \\
0.02 & 113 & 188 & 1 & 123 & 134 & 123 & 134 & 1 & 3.81 & 103 \\
0.05 & 101 & 188 & 0 & 105 & 134 & 105 & 134 & 0 & 3.84 & 75 \\
0.10 & 95 & 188 & 0 & 97 & 134 & 97 & 134 & 0 & 3.90 & 68 \\
0.20 & 93 & 188 & 0 & 92 & 134 & 92 & 134 & 0 & 3.90 & 64 \\
\bottomrule
\end{tabular}
\par\smallskip\raggedright\footnotesize Questions exclude mandatory slots; unsafe counts are pool-unsafe releases. Ask-all and the silent default release in every case, CGC requests 6.10 bundles per case at every tolerance, and robust-first CGC requests 6.84 and has no unsafe release.
\end{table}

\subsection{Screening saves references when questions prune readings}\label{sec:screening}

\begin{figure}[t]
\centering
\includegraphics[max width=\linewidth]{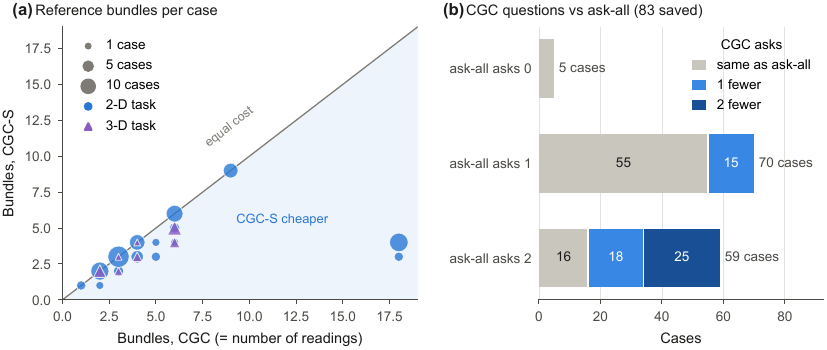}
\caption{Question and computation counts for 134 primary cases. (a) CGC-S
versus CGC reference bundles per case, with symbol size indicating
multiplicity. (b) Cases grouped by the number of non-mandatory questions
ask-all asks; each bar segment counts the cases in which CGC asks the same
number, one fewer, or two fewer. Requested
bundles are charged even when cached offline. The aggregate saving
depends on tasks with many readings. CGC: consequence-guided clarification; CGC-S: its screening variant.}
\label{fig:cost}
\end{figure}

CGC requests 818 two-start reference bundles and CGC-S 514 over the primary replay,
a 37.2\% case-weighted reduction (Fig.~\ref{fig:cost}). The reduction is
18.7\% when tasks are weighted equally and 9.9\% without the largest
reading set, L22. Thus the aggregate saving depends strongly on a task
with many readings. Screening saves references when answers remove
readings; harmless alternatives remain expensive because all their
references are requested before release.

Measured batch times put a median primary case at 155~s for CGC-S versus
36~s for ask-all. The runs shared processors, so these measurements
include contention. Relative to ask-all, each avoided question costs
about 4.6 extra two-start reference bundles, or 169~s in two-dimensional
tasks and 2,315~s in three-dimensional ones. These are optimization-only
sums. Cross-evaluation time and human answering time were not included,
so they do not establish an overall workflow break-even.

CGC-S requests 3,254 distinct design--reading cross-evaluations versus
8,420 full-table entries summed over cases. Support groups determine
how many factorizations these require. Controlled-track policy replays
show the same pruning mechanism (\supptab{19}); the full timing,
weighting, and screening-statistic comparisons are in \suppsec{14}.

\subsection{Richer reading sets make question order matter}\label{sec:union}

The union encoding admits 41 tasks and 240 readings. Ask-all asks 531
questions, CGC 336, and CGC-S 380. CGC releases in 232 cases and CGC-S
in 224 without a pool-unsafe release at $\tau=0.05$ (\supptab{7}).
Withholding occurs in tasks with four open slots under budget three;
a budget of four permits all releases in those tasks.

On this richer set, full CGC asks fewer questions and releases more often
than either the declared or reverse fixed order with the same full-table
check. Greedy selection can nevertheless use more questions than the best fixed slot order selected for each task after seeing all
replay outcomes; all order and budget results are in \suppsec{7}
and \suppsec{14}.

CGC-S requests 1,291 bundles versus 2,686 for CGC, a 51.9\% reduction,
but asks more questions and withholds more releases. The saving is
therefore not at equal decisions. At $\tau=0.02$, CGC-S makes three
pool-unsafe union releases (\supptab{12}). The reading set controls both
the cost and the scope of the observed release outcomes.

\section{Discussion}\label{sec:discussion}
\subsection{What consequence-guided clarification changes}
The acceptance test accounts for most of the primary-set question saving.
It asks whether a computed design already covers every open reading,
rather than requiring every descriptive detail to be settled. Checking
only the default design supplies 71 of the 83 avoided questions; the
other computed candidates supply the remaining 12. Question order adds
no saving on these mostly one-slot tasks. It matters on the richer union
encoding, but the greedy rule has no general question-optimality
guarantee. The evidence therefore supports the release check more
strongly than any particular question order.

This saving requires computation. Each avoided answer costs about 4.6
additional two-start reference bundles for CGC-S relative to ask-all on
the primary benchmark. That comparison excludes cross-evaluation and
human answering time. It does not establish an overall workflow
break-even. The relative costs of asking, screening, and attempting a
robust design determine which evaluated policy is cheapest; the detailed
cost crossings are in \suppsec{14}. Where answers are cheap and
immediate, ask-all is preferable. The bundle counts transfer to other
settings more directly than seconds measured on these grids and shared
processors.

An available answerer and literature reconstruction lead to different
uses. With slow replies or questions that require a specialist, the
check could decide which answers are worth obtaining. This regime was
not tested with real answerers. Without an answerer, no reply can prune
the readings, so the question rule has no role. The useful output is
then whether one computed design covers the listed alternatives.
Question savings also differ from interaction rounds: batching every
question can make ask-all use fewer rounds.

\subsection{Coverage and numerical limits}
A passed check covers only the enumerated readings. The omitted-intent
controls show that a design can pass throughout its pool and fail
severely just outside it (\suppsec{8}). The exploratory language-model
test also leaves the finite reading set in need of expert review
(\suppsec{6}). Neither minimax computation nor a language-model front
end certifies an intent that was never represented.

Computed references create a second boundary. Two starts and numerical
acceptance checks provide usable candidates, with no certificate of
global optimality. Better references push three L18 releases above the
primary tolerance, and mesh refinement changes decisions in three
selected tasks (Section~\ref{sec:robustness}). The two-start spread can
flag a numerically unresolved release, but it is a heuristic margin,
not an error bound. The conservative lower-bound replay releases far
fewer cases. CGC-S adds a local-reference approximation: its zero
exceedance count at $\tau=0.05$ does not persist at stricter tolerances.
Extreme losses can also depend on artificial void stiffness. These are
separate limits on optimization, discretization, and the decision check.

Sampling a continuous slot supplies no interval certificate. If a slot
enters the load vector affinely, a fixed design's compliance is a convex
quadratic whose maximum over an interval lies at an endpoint. The
reading-specific reference values are not bounded in the same way.
The midpoint and sub-sampling checks therefore test selected values,
with no guarantee for every value between them.

\subsection{Benchmark scope and provenance}\label{sec:provenance}
The benchmark is an arXiv reconstruction set used to develop the
policies, with no held-out or prospective test. AI screening and initial
encodings were followed by blind model re-annotation and author
adjudication; direct-source coverage differed between the authors.
Consensus and union encodings expose some of the resulting judgment.
Removing cases with weaker direct-source coverage, halving selected
position ranges, and changing load-introduction regions alters individual
outcomes while retaining frequent default failures (\suppsec{14}).
Most sampled consequences are either below tolerance or far above it;
a set with more graded consequences would test the check more severely.

Frozen preprints define the examples even when published versions
exist. Most published versions checked retained the open details, but
version and figure coverage was incomplete (\suppsec{9}). These counts
therefore do not estimate ambiguity in final journal articles.
Standardized reconstructions can also change source constraints, load
regularization, and material parameterization. Six of the seven
three-dimensional tasks were not refined, and four use grids coarser
than their sources. The tested grids do not establish industrial-scale
behavior. Exact answers leave response errors, question wording, and
operator time untested; no real engineering briefs were sampled.

\subsection{Implications for computational design workflows}
The release check places clarification within engineering design as
decision making \cite{Hazelrigg1998}. Value-of-information analysis
asks when further information is worth its cost \cite{Howard1966},
including simulation refinement \cite{Panchal2008} and clarification
questions \cite{Rao2018}. Requirements engineering studies ambiguity in
natural-language specifications \cite{BerryKamsties2004}. CGC connects
these views to a computed mechanical consequence: an unanswered detail
matters to acceptance when no available design covers its alternatives.

A practical interface could show the open readings, candidate designs,
proposed question, and basis of the acceptance test. Explicit passive
load and support regions would make reconstruction assumptions visible;
the matched controls show that these can dominate a positional
ambiguity, although they do not resolve every failure. Scaling remains
a constraint. With $n$ readings, full CGC requires $n$ reference bundles
and $n^2$ design--reading evaluations; the benchmark has at most 18
readings per task. We did not test a cheaper screen for larger sets.

Within these limits, the check makes a design decision inspectable: it
shows which enumerated specifications a computed design can cover and
which unanswered details prevent acceptance. With an exact answerer,
this can guide the choice to ask, compute, or commit. Without one, it
can still test whether a single computed design covers the listed
alternatives. The measured optimization costs let a future workflow
compare that benefit with the cost of obtaining answers.

\section{Conclusion}\label{sec:conclusion}
Silent defaults failed the 5\% loss tolerance at another reading, by
excess loss or infeasibility, in 26 of 36 encoded literature tasks. The
acceptance test reduced non-mandatory questions from 188 for ask-all to
105 for CGC and CGC-S. Checking only the default design required 117,
so most of the saving came from the check. Minimax ordering added no
primary-set saving over the declared slot order; it mattered on richer
reading sets.

These outcomes depend on the finite readings and computed references.
Better references pushed three accepted L18 cases just above tolerance,
and mesh refinement changed decisions in three selected tasks. The
lower-bound replay certified far fewer releases. Most decisions are
therefore comparisons with computed designs, with no certificate against
the unknown optimum.

Screening reduced reference requests from 818 to 514, but each avoided
question still cost about 4.6 additional two-start reference bundles
relative to ask-all. Cross-evaluation and human answering time were
excluded, so this is an optimization cost comparison. The tasks were
also used to develop the policies. Within that scope, the check shows
which questions the evaluated policies can avoid and what computation
they require. With no answerer, it instead tests whether one computed
design covers the listed alternatives.

\section*{Supplementary material}
The Supplementary Information, supplied with this preprint as an
ancillary file, contains the screening protocol,
all task encodings and exclusions, adjudication and re-annotation details,
the language-model front-end experiment with its prompts and scoring, and
numerical controls. The two original
author adjudication records are supplied as Supplementary Files A and B;
their scope and reconciliation are explained in \suppsec{3}.
\section*{Data and code availability}
The study's data comprise the frozen literature frame, screening and
adjudication records, task specifications, optimization results, policy
replays, and figure and table sources. The computational package contains
the solver and experiment scripts needed to regenerate the reported
results from the stored optimization records. Data and code will be
released at \url{https://github.com/nbbllxx0/consequence-guided-clarification}
when the preprint is posted, with an archival identifier assigned at
release. Third-party paper texts and the copies supplied to the
language-model experiments are excluded from the public release;
the Supplementary Information identifies their public arXiv sources.
\section*{CRediT authorship contribution statement}
Shaoliang Yang: Conceptualization, Methodology, Software, Validation,
Formal analysis, Investigation, Data curation, Visualization,
Writing -- original draft, Writing -- review \& editing.
Jun Wang: Conceptualization, Methodology, Validation, Data curation,
Supervision, Funding acquisition, Writing -- review \& editing.
\section*{Declaration of competing interest}
The authors declare that they have no known competing financial interests
or personal relationships that could have appeared to influence the work
reported in this paper.
\section*{Funding}
This work was supported by Santa Clara University.
\section*{Declaration of generative AI and AI-assisted technologies in the manuscript preparation process}

Language models used as research components (literature screening, first
task encodings, blind re-annotation, and the front-end experiment of
\suppsec{6}) are described in the paper. The authors conceived the study,
made every protocol and design decision, supervised each model run and
collected data, conducted the analysis and wrote the manuscript, and are
responsible for all conclusions. Under their direction, Claude Code
(Anthropic, running Claude Opus~5.5) helped implement the solver,
experiment, and plotting scripts; it and OpenAI Codex also edited and checked manuscript text. The
authors reviewed and edited all of this work and take full responsibility
for the content of the published article. No AI system is an author.

\bibliographystyle{elsarticle-num}
\bibliography{references}
\end{document}